\documentclass[aps,letterpaper,10pt,prc,tightenlines,floatfix,superscriptaddress,nofootinbib]{revtex4}
\usepackage[bookmarks,dvips,pdfhighlight=/O,pdfborder={0 0 0},pdfstartview=FitH]{hyperref}

\usepackage{graphicx}
\usepackage{amsmath,amssymb,amsopn,bm}
\usepackage{color}

\newcommand{\eg}{\textit{e.g.}}
\newcommand{\ie}{\textit{i.e.}}
\newcommand{\lsim}{\lesssim}
\newcommand{\gsim}{\gtrsim}
\newcommand{\dd}{\mathrm{d}}
\newcommand{\erw}[1]{\left \langle #1 \right \rangle}

\newcommand{\MeV}{\, \mathrm{MeV}}
\newcommand{\GeV}{\, \mathrm{GeV}}
\newcommand{\fm}{\, \mathrm{fm}}

\begin{document}

\title{Resonance Recombination Model and Quark Distribution Functions
  in the Quark-Gluon Plasma} 

\author{L. Ravagli}
\author{H. van Hees}
\altaffiliation{present address: Institut f{\"u}r Theoretische
 Physik, Justus-Liebig-Universit{\"a}t Giessen, Heinrich-Buff-Ring 16,
 D-35392 Giessen, Germany}
\author{R. Rapp} 
\affiliation{Cyclotron Institute and Physics Department, Texas A\&M
  University, College Station, Texas 77843-3366, U.S.A.}

\date{May 14, 2009}

\begin{abstract} 
  We investigate the consequences of space-momentum correlations in
  quark phase-space distributions for coalescence processes at the
  hadronization transition. Thus far it has been proved difficult to
  reconcile such correlations with the empirically observed constituent
  quark number scaling (CQNS) at the Relativistic Heavy-Ion Collider
  (RHIC). To address this problem we combine our earlier developed quark
  recombination model with quark phase-space distributions computed from
  relativistic Langevin simulations in an expanding Quark-Gluon Plasma
  (QGP). Hadronization is based on resonance formation within a
  Boltzmann equation which recovers thermal equilibrium and obeys energy
  conservation in the quark-coalescence process, while the fireball
  background is adjusted to hydrodynamic simulations of semi-central
  Au-Au collisions at RHIC. To facilitate the applicability of the
  Langevin process, we focus on strange and charm quarks. Their
  interactions in the QGP are modeled using leading-order perturbative
  QCD augmented by effective Lagrangians with resonances which smoothly
  merge into hadronic states formed at $T_c$. The interaction strength
  is adjusted to reproduce the empirical saturation value for the
  quark-elliptic flow, $v_{2,q}^{\mathrm{sat}}\simeq 7$-$8\%$. The
  resulting $\phi$ and $J/\psi$ elliptic flow recover CQNS over a large
  range in transverse momentum ($p_T$) within a few percent. As a
  function of transverse kinetic energy, both the quark spectra from the
  Langevin simulations and the meson spectra generated via resonance
  recombination recover CQNS from zero to at least $3\GeV$.
\end{abstract}

\maketitle

\section{Introduction}
\label{sec_intro}

The mechanism of hadronization, \ie, the conversion of quarks and gluons
produced in hadronic or electromagnetic reactions into colorless
hadrons, is a non-perturbative problem that is presently not calculable
within the theory of strong interactions, Quantum Chromodynamics
(QCD). For partons produced at large transverse momentum, $p_t$, the
factorization theorem of QCD allows to treat the hadronization process
via a so-called fragmentation function, which is universal and can, in
principle, be determined empirically. At low momentum this scheme is no
longer applicable and other hadronization mechanisms become
relevant. The quark coalescence model (QCM) has provided a
phenomenologically successful framework to understand several
non-perturbative features of hadron production in hadronic
collisions. In elementary ($p$-$N$ and $\pi$-$N$) reactions, flavor
asymmetries in kaon and charmed hadron spectra have been associated with
the recombination of produced strange and charm quarks with valence (and
sea) quarks in target and
projectile~\cite{Das:1977cp,Braaten:2002yt,Rapp:2003wn}. In heavy-ion
reactions at the Super Proton Synchrotron (SPS)~\cite{Biro:1994mp} and
the Relativistic Heavy-Ion Collider (RHIC), recombination of quarks from
a thermalized Quark-Gluon Plasma
(QGP)~\cite{Hwa:2002tu,Greco:2003mm,Fries:2003kq,Molnar:2003ff,Lin:2004en,
  Zimanyi:2005nn,Miao:2007cm,Ravagli:2007xx,Ayala:2007cp,Cassing:2008sv}
gives a simple and intuitive explanation of several unexpected features
in the observed hadron spectra, most notably the large baryon-to-meson
ratio and the rather universal constituent quark number scaling (CQNS)
of the elliptic flow coefficient, $v_{2,h}(p_T) \equiv n_q
v_{2,q}(p_T/n_q)$ ($n_q$: number of valence quarks in hadron $h$, $p_T$:
transverse momentum of $h$). This scaling relation implies the momenta
of the coalescing quarks to be collinear, and its implementation in QCMs
usually restricts their applicability to sufficiently large momenta so
that the associated non-conservation of energy in the hadron formation
process is small. Since at high $p_T$ parton fragmentation is expected
to take over, the typical range of applicability of QCMs is at
intermediate momenta, $2\;\mathrm{GeV} \lsim p_T \lsim 6\;\mathrm{GeV}$.
In our recent work~\cite{Ravagli:2007xx}, we have suggested a
reinterpretation of quark coalescence in terms of hadronic resonance
formation, implemented via $q+\bar q \to M$ scattering into a Boltzmann
equation ($M$: meson). Energy conservation is obeyed by utilizing
hadronic reaction rates, along with detailed balance, based on pertinent
spectral functions. In addition, we have shown that this approach
correctly recovers the thermal equilibrium limit, which enabled a more
controlled extension of the coalescence mechanism to low $p_T$ and to
make contact with the phenomenologically successful hydrodynamic
description of bulk matter at RHIC.

Another aspect that has evaded a satisfactory explanation in QCMs at
RHIC is the question of space-momentum correlations in the underlying
(thermal) quark distribution functions (see, e.g.,
Ref.~\cite{Fries:2008hs} for a recent critical review). In hydrodynamic
models the elliptic flow of produced hadrons is a collective effect that
implies a definite correlation between the particle's momentum and its
spatial position in the fireball, \ie, a (locally thermalized) fluid
cell moving into a specific direction preferentially emits hadrons in
that same direction. Such a correlation is neglected within the
so-called ``factorized'' implementation of the parton $v_{2,q}(p_T)$
which does not carry any spatial dependence (and is therefore identical
regardless of the quark's position inside the fireball). While this
approximation straightforwardly recovers the empirical constituent-quark
number scaling (CQNS) of the hadron elliptic flow, it is at variance
with the hydrodynamic description of $v_2$ as a collective expansion
effect.
 
Previous attempts to incorporate space-momentum correlations into QCMs
have found the empirically observed CQNS to be rather
fragile~\cite{Pratt:2004zq,Molnar:2004rr,Greco:2005jk}. Part of the
problem is the construction of a realistic transition from the thermal
to the kinetic regime of the underlying parton phase-space distribution
functions, as characterized by the ``saturation'' (leveling off) of the
empirical parton $v_2$ at about $p_t\simeq 1\GeV$. In
Ref.~\cite{Pratt:2004zq} several elliptic ``deformations'' of a thermal
blast-wave parameterization have been considered, motivated by different
plausible realizations of $v_2$. While some features could be ruled out
being incompatible with the empirical CQNS, other assumptions did not
spoil the latter. In Ref.~\cite{Greco:2005jk} a reduction of the boost
velocity at higher momenta was introduced, entailing a violation of CQNS
at the 20\% level. It therefore seems that purely phenomenological
prescriptions of space-momentum correlations and associated $v_2$ did
not arrive at a conclusive interpretation of the key features underlying
the quark distribution functions. In Ref.~\cite{Molnar:2004rr}, based on
numerical transport simulations, it was even argued that rather delicate
cancellations must be at work to obtain CQNS for the thermal components,
thus raising doubts on the robustness of the coalescence approach. In
view of the broad empirical applicability of CQNS across different
centralities, system sizes and collision
energies~\cite{Adare:2006ti,Abelev:2007qg}, such an interpretation would
be difficult to reconcile with experiment.

In the present paper we adopt a microscopic approach to compute quark
distributions in four-dimensional phase space (transverse position and
momentum) by employing relativistic Langevin simulations for strange and
charm quarks within an expanding thermal QGP background. In a strict
sense, the underlying Fokker-Planck equation is applicable for a
diffusive treatment of heavy and/or high-momentum quarks, \ie, in a
regime where the momentum transfers from the heat bath are small. Our
simulations for low-momentum ($p_t\lsim 1\,\mathrm{GeV}$) strange quarks
are thus at the boundary of applicability of a Fokker-Planck treatment
and may be considered as extrapolations thereof. The Langevin approach
has the attractive feature that it naturally encodes the transition from
a thermal to a kinetic regime. In particular, when simulating
heavy-quark (HQ) motion in an expanding QGP fireball for non-central
collisions, this transition reflects itself in a saturation of the
elliptic flow~\cite{vanHees:2005wb,Moore:2004tg}, a key ingredient to
CQNS in light-hadron spectra observed at RHIC. It turns out that
Langevin simulations preserve the $v_2$ saturation feature when applied
to strange quarks (with thermal masses of $\sim$$0.5\,\mathrm{GeV}$).
We will therefore investigate whether the resulting quark distribution
functions, evolved to the hadronization transition and injected into our
resonance recombination approach, allow for a better (microscopic)
understanding of space-momentum correlations in the coalescence
process. In view of the rather delicate dependence of CQNS on these
correlations (as discussed above), a realistic treatment of the
kinematics in the hadron formation process is mandatory, including
energy-momentum conservation, non-collinear kinematics, and a well
defined equilibrium limit. The recombination approach developed in
Ref.~\cite{Ravagli:2007xx} satisfies these requirements. In addition to
this, the QGP evolution and subsequent hadronization are linked via the
ansatz that resonances play an essential role in hot QCD matter around
$T_c$. This scenario is consistent with effective potential models where
a non-perturbative description of the strongly coupled QGP (sQGP) is
realized via bound~\cite{Shuryak:2004tx} and/or
resonance~\cite{Mannarelli:2005pz,vanHees:2007me} states of deconfined
partons. Recent lattice QCD computations support the picture of various
(light and strange) hadronic states surviving up to temperatures of
$\sim$1.5-2~$T_c$~\cite{Karsch:2003jg,Asakawa:2003nw}.

Our article is organized as follows. In Sec.~\ref{sec_boltz} we review
our earlier developed model~\cite{Ravagli:2007xx} for resonance
hadronization based on the Boltzmann equation. In Sec.~\ref{sec_parton}
we elaborate the computation of the phase-space distributions of quarks
obtained from Langevin simulations of an expanding QGP fireball at RHIC.
In Sec.~\ref{sec_meson} we discuss the numerical results for the $v_2$
coefficients of $\phi$ and $J/\psi$ mesons within our model, and discuss
their properties in terms of CQNS in both transverse momentum and
transverse kinetic energy, $K_T$. Sec.~\ref{sec_concl} contains our
conclusions.

\section{Recombination from the Boltzmann Equation}
\label{sec_boltz}

Following Ref.~\cite{Ravagli:2007xx}, our description of hadronization
at the critical temperature, $T_c$, is based on the Boltzmann equation
using resonance quark-antiquark cross sections to compute meson spectra
in terms of underlying anti-/quark phase-space distributions,
$f_{q,\bar{q}}$ (baryons could be treated in a similar way, \eg, in a
two-step process using subsequent quark-quark and quark-diquark
interactions; in the present paper, we will focus on mesons). The meson
phase-space distribution, $F_M$, is determined by the equation
\begin{equation}
\label{boltz}
\left( \frac{\partial}{\partial t} 
+\vec{v}\cdot\vec{\nabla} \right) F_M(t,\vec x,\vec p) 
=-\frac{\Gamma}{\gamma_p} \, 
F_M(t,\vec x,\vec p)+\beta(\vec x,\vec p) \ , 
\end{equation}
where $\vec p$ and $\vec x$ denote three-momentum and position of the
meson, $M$, and $\vec{v}=\vec p/E_M(p)$ ($m$, $E_M(p)$=$\sqrt{m^2+\vec
  p^2}$: meson mass and energy). The total meson width, $\Gamma$, is
assumed to be saturated by the coupling to quark-antiquark states,
$M\leftrightharpoons q + \bar q$ and taken to be constant, with the factor
$\gamma_p=E_M(p)/m$ accounting for Lorentz time dilation, see also
Ref.~\cite{Miao:2007cm}. Integrating over the fireball volume leads to
the momentum-distribution function of the meson, $f_M$, and the
pertinent transport equation
\begin{equation}
\label{boltz-mom}
f_M(t,\vec p)=\int \dd^3 x \, F_M(t,\vec x,\vec p), \quad 
\frac{\partial}{\partial t} f_M(t,\vec p) 
=-\frac{\Gamma}{\gamma_p} \, f_M(t,\vec p)+g(\vec p).
\end{equation}
The drift term vanishes upon integration over $\vec x$ since it is a
total divergence: $\vec{v} \cdot \vec{\nabla} f_M(t,\vec x,\vec
p)=\vec{\nabla} \cdot [\vec{v} f_M(t,\vec x,\vec p)]$. The relation of
the gain term, $g(\vec p)$, to the underlying microscopic interaction is
given by
\begin{equation}
\label{gain}
g(\vec p)=\int \dd^3 x \beta(\vec x,\vec p)= \int 
\frac{\dd^3 p_1 \dd^3 p_2}{(2 \pi)^6} \int \dd^3 x 
\ f_q(\vec x,\vec p_1) \ f_{\bar{q}}(\vec x,\vec p_2) \ \sigma(s) \ 
v_{\mathrm{rel}}(\vec p_1,\vec p_2) \ \delta^{(3)}(\vec p-\vec p_1-\vec p_2)
\end{equation}
with $\sigma(s)$ the cross section for the process $q+\bar{q}\to M$ at
center-of-mass (CM) energy squared, $s=(p_1^{(4)}+p_2^{(4)})^2$, where
$p_{1,2}^{(4)}$ are the four-momenta of quark and antiquark. The quark
phase-space distribution functions are normalized as $N_{q,\bar{q}}=\int
\frac{\dd^3 x \;\dd^3 p}{(2\pi)^3} f_{q,\bar{q}}(\vec x,\vec p)$.
Throughout this paper, quarks will be assumed to be zero-width
quasi-particles with an effective mass $m_q$ (which contains both
thermal and bare contributions). The classical nature of the Boltzmann
equation warrants the use of classical distribution functions for all
the particles, and we assume zero chemical potentials for all quark
species. The cross section is approximated by a relativistic
Breit-Wigner form,
\begin{equation}
\label{cross}
\sigma(s)=g_{\sigma}\frac{4\pi}{k^2}
\frac{(\Gamma m)^2}{(s-m^2)^2+(\Gamma m)^2} \ ,
\end{equation}
where $g_{\sigma}=g_M/(g_q g_{\bar{q}})$ is a statistical weight given
in terms of the spin (-color) degeneracy, $g_M$ ($g_{q,\bar{q}}$), of
the meson (anti-/quark), and $k$ denotes the quark three-momentum in the
CM frame. With $M\leftrightharpoons q +\bar q$ being the only channel,
it follows that
$\Gamma_{\mathrm{in}}=\Gamma_{\mathrm{out}}=\Gamma$. Detailed balance
requires the same $\Gamma$ in the loss term on the right-hand side of
Eq.~(\ref{boltz}), thus ensuring the correct equilibrium limit with
$\tau=1/\Gamma$ the pertinent relaxation time. This formulation
conserves four-momentum and applies to all resonances $M$ with masses
above the $q\bar q$ threshold, \ie, for a positive $Q$ value,
\begin{equation}
\label{masscondition}
Q= m - (m_q+m_{\bar q}) \gsim 0 . 
\end{equation}
If the $2\rightarrow 1$ channel proceeds too far off-shell, \ie, $Q<0$
and $\Gamma < |Q|$ (\eg, for pions), other processes need to be
considered, \eg, $q+\bar{q}\rightarrow M+g$ (which, in principle, is
possible in the present framework by implementing the respective cross
sections). We note that the majority of the observed pions are believed
to emanate from resonance decays ($\rho$, $\Delta$, $a_1$ etc.); in
addition, hydrodynamic calculations suggest that the elliptic flow in
heavy-ion collisions at RHIC does not change much after
hadronization~\cite{Kolb:2003dz}. We also note that in the absence of a
confining interaction individual quarks and antiquarks remain a part of
the heat bath.

\begin{figure}[!t]
\begin{center}
\includegraphics[width=0.4\textwidth]{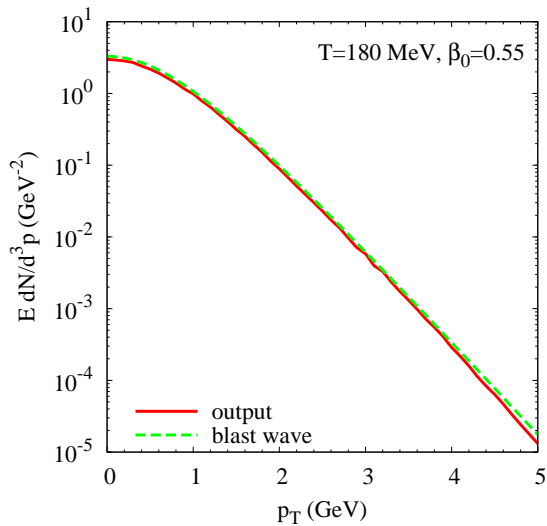}
\end{center}
\caption{(Color online) $p_T$ spectra for $\phi$ mesons using the
  Boltzmann recombination equation in the equilibrium limit,
  Eq.~(\ref{eqboltzappr}), based on blast-wave input distributions for
  strange quarks (solid line), compared to $\phi$ spectra directly
  obtained from a blast-wave expression with the same fireball
  parameters (temperature $T=180 \MeV$, radial expansion surface
  velocity $\beta_0=0.55c$). The $\phi$-meson resonance parameters are
  $m_{\phi}=1.02 \GeV$, $\Gamma_\phi=50 \MeV$, and the $s$-quark mass is
  $m_s=0.45 \GeV$.}
\label{ptequil}
\end{figure}
The equilibrium limit is readily recovered within our approach by
imposing the stationarity condition,
\begin{equation}
\label{eqcond}
\frac{\partial}{\partial t} f_M(t,\vec p)=0 \ .
\end{equation}
Then (\ref{boltz-mom}) is immediately solved by
\begin{equation}
\label{eqboltzappr}
f_M^{\mathrm{eq}}(\vec p)=\frac{\gamma_p}{\Gamma} g(\vec p) \ ,
\end{equation}
which represents the large time limit of the Boltzmann equation and is
the expression that comes closest to the conventional QCM approximation.
For hadronization times less or comparable to the relaxation time,
$\tau$, the equilibrium limit will not be reached and an explicitly
time-dependent solution is in order. We have verified numerically that
Eq.~(\ref{eqboltzappr}) accurately recovers the standard thermal
Boltzmann distribution for a meson $M$ at temperature $T$, if the
constraint of a positive $Q$ value is satisfied (for negative $Q$ the $2
\to 1$ channel is inoperative), as illustrated in Fig.~\ref{ptequil} for
the case of the $\phi$ meson: when using thermal quark input
distributions with radial flow (blast-wave model), the computed
equilibrium expression (\ref{eqboltzappr}) is in excellent agreement
with the same blast-wave expression (identical temperature and flow
profile) directly applied at the meson level. This reiterates the close
connection between equilibration and energy conservation in the approach
of Ref.~\cite{Ravagli:2007xx}, providing a significant improvement of
previous QCMs.

\section{Partonic Spectrum}
\label{sec_parton}

While the blast-wave model provides a convenient description of the
thermal component of empirical hadron (and/or parton) spectra, the
transition to the kinetic, and eventually hard-scattering, regime is
more involved, especially with regard to phase-space correlations within
QCMs, as discussed in the Introduction. In an attempt to generate
realistic quark input distributions for meson formation processes in the
vicinity of $T_c$, we here adopt a Fokker-Planck approach for test
particles evolving in a thermally expanding QGP background. The latter
is parameterized with guidance from hydrodynamic models for central and
semicentral Au-Au collisions at RHIC, implementing empirical values
(\ie, adjusted to experiment) for bulk matter properties such as total
entropy, radial and elliptic flow. The elliptic flow is parameterized in
terms of a flow profile, whose direction at each transverse position is
perpendicular to confocal elliptic isobars. Its magnitude is chosen to
increase linearly in distance from the center with an average boundary
value of $\beta_0=0.55c$ at the end of the mixed phase of the fireball
evolution, \ie, at the hadronization time. The acceleration is adjusted
so that the fireball at hadronization is approximately circular, with
bulk $v_{2,q} \simeq 5.5\%$. This model has been employed before in the
context of heavy-quark (HQ), \ie, charm and bottom spectra at
RHIC~\cite{vanHees:2005wb}, and good agreement with hydrodynamic
simulations has been found~\cite{Moore:2004tg} for the same HQ diffusion
coefficient. HQ observables, which at present are semileptonic
single-electron decay spectra~\cite{Adare:2006nq,Abelev:2006db}, exhibit
an unexpectedly large suppression and elliptic flow which cannot be
understood within perturbative QCD (pQCD) including both radiative and
elastic scattering.  The key microscopic ingredient in
Ref.~\cite{vanHees:2005wb} are resonant heavy-light quark interactions
mediated via effective (broad) $D$ and $B$ mesons in the
QGP~\cite{vanHees:2004gq}, inspired by the findings of thermal lattice
QCD. In connection with a ``conventional'' coalescence
afterburner~\cite{Greco:2003vf} at $T_c$, the predictions for
single-electron suppression and $v_2$ turned out to be in fair agreement
with data~\cite{Adare:2006nq,Abelev:2006db}. In more recent
work~\cite{vanHees:2007me}, the effective interactions have been
replaced by in-medium $T$-matrices based on finite-temperature HQ
potentials extracted from lattice QCD; these calculations not only
confirmed the interaction strength generated by heavy-light resonances
in an essentially parameter-free way, but also identified pre-hadronic
meson and diquark channels as the most relevant ones. This, in turn,
provides a direct link between two main discoveries at RHIC, namely the
strongly interacting nature of the sQGP and quark coalescence from a
collective partonic source.
  
In the present paper, we build upon the above findings by extending the
Fokker-Planck approach to strange ($s$) quarks. While its applicability
criterion, $m_t\gg q \sim T$ ($m_t=\sqrt{m^2+p_t^2}$: transverse mass,
$q$: momentum transfer in a typical scattering), seems to be only
marginally satisfied for momenta $p_t\lsim 1\GeV$ (at least in the early
phases of the QGP evolution), we note that most of the fireball
evolution occurs for temperatures close to $T_c$. At higher $p_t$ (or
$m_t$) one enters the kinetic regime where the Fokker-Planck treatment
becomes reliable again. With these limits properly satisfied, one may
hope that the Langevin simulations also accomplish a reasonable
description of the low-$p_t$ regime ($p_t \lsim 1\,\mathrm{GeV}$) for
strange quarks, including realistic space-momentum correlations, which
is one of the main objectives in our work. It remains to specify the
interaction strength of the strange- and charm-quark species. Here we
take further guidance from phenomenology by requiring that the final
quark $v_2(p_t)$ exhibits the characteristic saturation (or maximum)
value of $\sim$$7.5\%$. Our baseline interaction for the stochastic
Langevin force is elastic pQCD scattering with a rather large value of
$\alpha_s=0.4$ (which can be thought of as containing radiative and/or
parts of non-perturbative contributions). However, additional
non-perturbative interactions are necessary to achieve a sufficiently
large $v_2$. As in Refs.~\cite{vanHees:2004gq,vanHees:2005wb} we
associate these with mesonic resonance states with an interaction
strength controlled by the resonance width (with a larger width implying
stronger coupling). For $s$ quarks ($m_s=0.45\GeV$) the ``heavy-light''
resonances require a width of $\Gamma_{s\bar q} \simeq 0.3\GeV$, and
$\Gamma_{c\bar q} \simeq 0.6\GeV$ for $c$ quarks ($m_c=1.5\GeV$), which
is compatible with
Refs.~\cite{vanHees:2004gq,vanHees:2005wb}\footnote{We
  recall~\cite{Ravagli:2007xx} that the only requirement on the quark
  and meson masses is that the latter are above the two-quark threshold;
  within this restriction variations in the mass and (positive) $Q$
  values have little effect on the recombination process. In fact, as we
  will see below, CQNS scaling emerges approximately independent of
  quark mass.}.  This hierarchy is qualitatively consistent with the
general expectation that resonance/bound-state formation is suppressed
with decreasing constituent mass (and also borne out of the microscopic
$T$-matrix calculations for $c$ and $b$ quarks in
Ref.~\cite{vanHees:2007me}). Finally, we have to specify the initial
quark distributions. For $c$ quarks we use the initial spectra as
constructed in Ref.~\cite{vanHees:2005wb}, as to reproduce $D$-meson and
semileptonic electron spectra in $p$-$p$ and d-Au collisions.  A similar
procedure is adopted for strange quarks: we parameterize the quark
spectra as a superposition of exponential and power-law spectra in a way
that experimental kaon spectra in $200\,\mathrm{GeV}$ $p$-$p$ collisions
are properly reproduced (using $\delta$-function fragmentation into
kaons at half the parent-quark momentum; as usual in QCMs, the role of
gluons is suppressed). In $AA$ collisions, the exponential ``soft'' part
is then scaled with the number of participants, $N_{\mathrm{part}}$, and
the power-law ``hard'' component with the number of collisions,
$N_{\mathrm{coll}}$, for a given centrality. With the interaction
strengths and initial conditions fixed, the quark phase-space
distribution in semi-/central Au-Au collisions are predicted from the
Langevin simulations at the end of the QGP (mixed) phase without further
adjustments, and serve as an input for the meson formation processes as
described in the previous Section. The framework developed here, \ie,
QGP evolution with resonance rescattering and recombination at $T_c$,
will be referred to as a ``Resonance Recombination Model'' (RRM).

The quark phase-space distributions resulting from the Langevin approach
embody strong correlations between spatial and momentum
variables. \textit{E.g.}, quarks at high $p_t$ tend to be located in the
outer layers of the fireball with a preferential alignment of the
momentum and position vector directions (\ie, the quark momentum tends
to point ``outward''). Likewise, the collective (radial and elliptic)
flow, which implies a well-defined (hydro-like) correlation between the
position of the fluid cell and its radial motion, imprints this
correlation on the (partially) thermalized components of the
Langevin-generated quark spectra. We recall again that the often used
factorized implementation of the $v_2$ coefficient in coalescence models
completely ignores these rather elementary dependencies.

The proper implementation of the differential phase-space information,
carried by the quark distribution functions (which is essential for a
realistic discussion of hadronic $v_2$-scaling properties), into the
hadronization formalism requires a few technical remarks. Since
thermalization in the longitudinal direction is somewhat controversial,
we assume the quark distributions to be homogeneous in the spatial
$z$-coordinate and flat in rapidity. This leaves four independent
transverse variables for each particle, which we choose in azimuthal
form, $(p_t,\phi_p,r_t,\phi_r)$, corresponding to the distribution ${\dd
  N_q}/{\dd^2 p_t \; \dd^2 r_t}$. This 4-D phase-space is then divided
into finite bins (with a maximum value of $p_t^{\mathrm{max}} \simeq 5
\;\mathrm{GeV}$). For each simulated test quark, its final location and
momentum is sorted into this grid. To warrant a (statistically)
sufficiently smooth behavior of the computed meson observables, a sample
of $\sim$$10^8$ test particles is needed. Finally, an interpolation
algorithm has been devised for converting the discretized distribution
back into a continuous function, to be plugged into the hadronization
formula. The algorithm recovers the periodicity properties of the two
angular variables, and converges to an arbitrary sampled function in the
limit of a large number of grid points. A suitable grid dimension
corresponding to the above variables amounts to, \eg, $(13, 96, 14,
12)$, where the large number of points in $\phi_p$ is dictated by the
large sensitivity in the determination of the elliptic flow coefficient,
$v_2(p_t)$.

We furthermore have to specify how to treat partons that escape the
fireball prematurely, \ie, before the end of the QGP/mixed phase is
reached. Clearly, these partons preferentially carry a high $p_t$, and,
upon exiting the fireball, could undergo a hadronization mechanism
different from coalescence, such as fragmentation. Since our fireball is
isotropic, the transition from the QGP to the vacuum is a sharp one and
it would be unrealistic to coalesce the exiting quark with a thermal
distribution at a temperature above $T_c$, or at a time much later than
the exit time (when the fireball has cooled down to $T_c$). We therefore
decide to include only partons in our hadronization framework which
remain inside the fireball throughout the entire QGP evolution. This
leads to an underestimation of the high-momentum part of the hadronic
spectra. A comprehensive calculation for quantitative comparison to
experiment also at high $p_t$ would require the treatment of the exiting
partons (hadronized with either fragmentation or
coalescence). Similarly, the hadronic $v_2$ we compute only reflects the
partons within the QGP fireball at the end of its lifetime (which,
however, do include non-thermal components from the Langevin simulation,
in addition to the thermalized part of the spectrum).

The generated spectrum, which originally represents a probability
distribution, requires a suitable normalization. Since the empirical
light and strange hadron spectra are consistent with chemical
equilibrium close to the expected phase boundary, we assume this to
apply at the quark level as well, at the critical temperature $T_c=180
\MeV$ of our fireball evolution. The fireball volume has been adjusted
to match the total entropy of the fireball to the experimental hadronic
final state multiplicities at $T=180\MeV$ at given collision centrality,
\eg, $V_{\mathrm{FB}}\simeq1200\fm^3$ for semicentral Au-Au collisions
(note that one fireball covers approximately $1.8$ units in
rapidity). For charm quarks we augment the chemical equilibrium number
by a fugacity factor, $\gamma_c \simeq 5$~\cite{Grandchamp:2003uw}, to
match their number to the expected hard production in primordial
nucleon-nucleon collisions (binary collision scaling; in
Ref.~\cite{Ravagli:2007xx} $\gamma_c\simeq 8$ at $T_c=170 \MeV$ leads to
the same number of $c\bar c$ pairs in the fireball). We note, however,
that the overall normalization has little impact on our main
considerations of space-momentum correlations and $v_2$ systematics.

Let us finally specify the assumptions on the rapidity ($y$)
distributions. As mentioned above, for both charm and strange quarks we
employ a step function as
\begin{equation}
  \frac{\dd N}{\dd y}=\left . \frac{\dd N}{\dd y} \right|_{y=0}
  \theta \left ( \frac{\Delta y}{2}-|y| \right) \ , 
\end{equation}
where the parameter $\Delta y$ depends on the quark mass and has been
adjusted to recover approximately the full-width-half-maximum of a
thermal $y$ spectrum (amounting to $\Delta y(s)=1.3$ and $\Delta
y(c)=0.8$ in connection with the quark masses quoted above).

\begin{figure}[!t]
\includegraphics[width=0.4\textwidth]{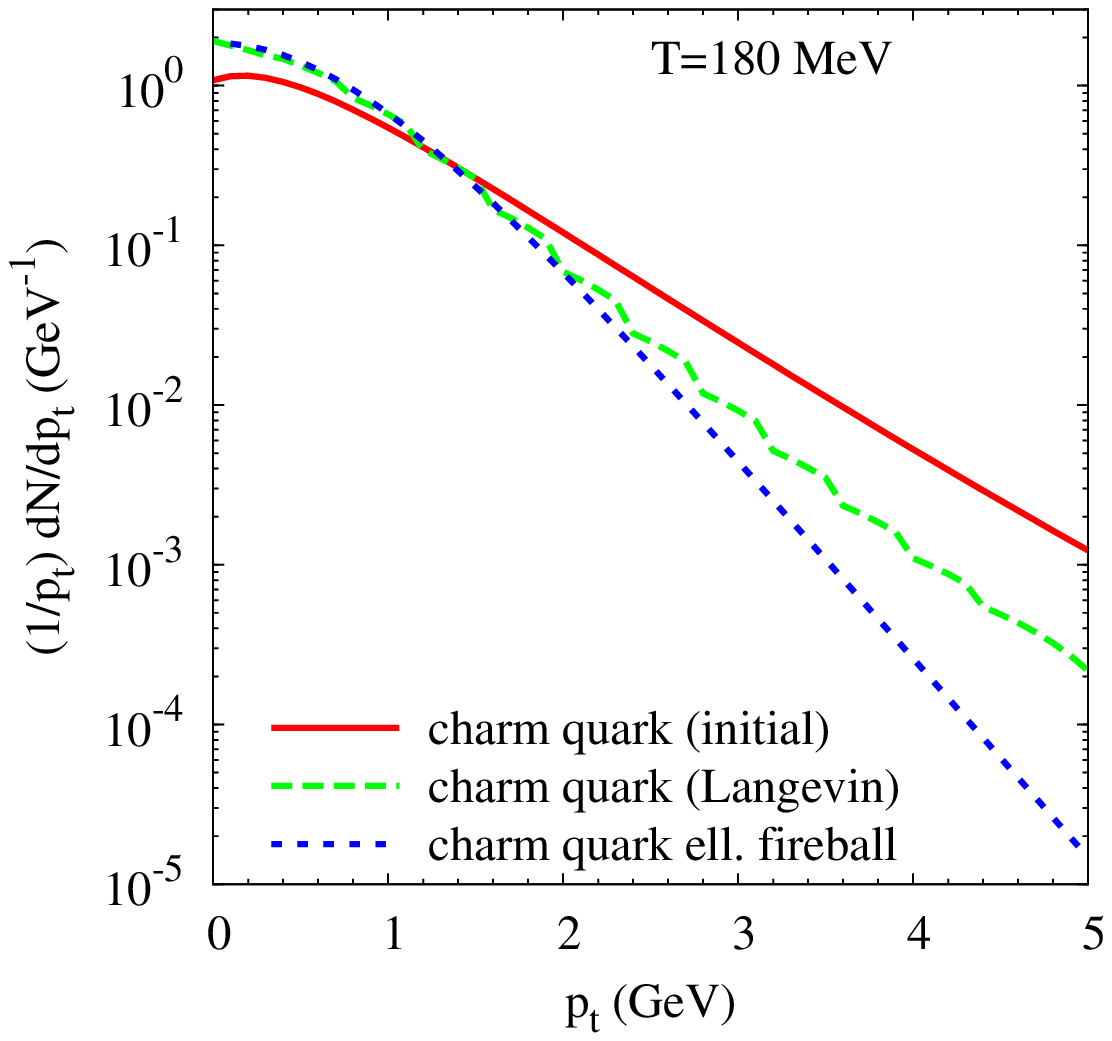}
\hspace{1.0cm}
\includegraphics[width=0.4\textwidth]{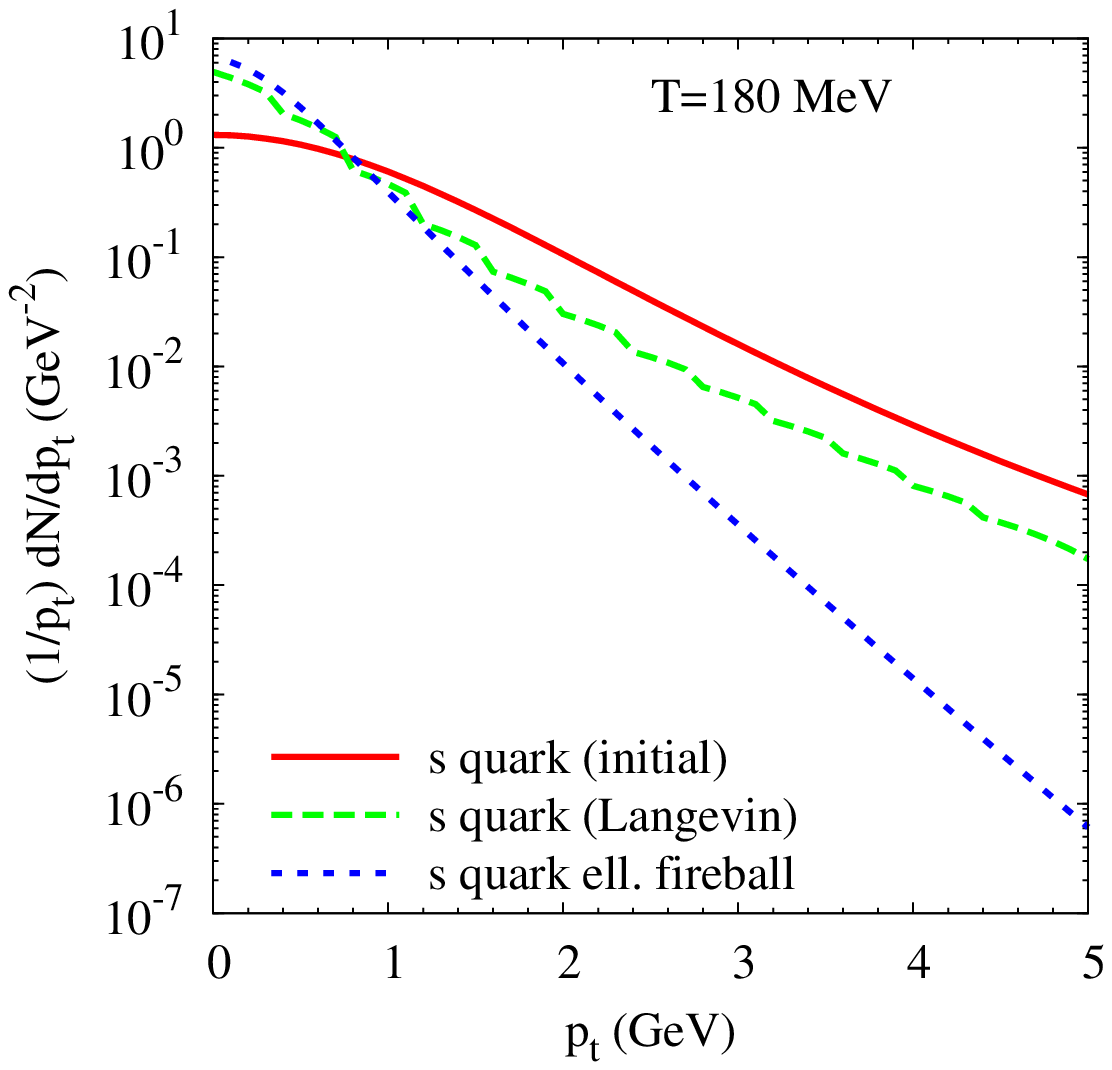}
\caption{(Color online) Quark $p_t$ distributions resulting from
  relativistic Langevin simulations of an expanding elliptic QGP
  fireball at the end of a QGP (mixed) phase at a temperature of
  $T=180\MeV$ and average radial surface expansion velocity,
  $\erw{\beta_0} \simeq 0.48c$. The numerically computed spectra
  (long-dashed lines) are compared to the initial spectra (solid lines)
  and to a blast-wave parameterization for particles with the same mass
  and the same fireball conditions ($T=180\MeV$ and the same flow field
  as used for the background medium in the Langevin simulation;
  short-dashed lines).  Left and right panels correspond to charm
  ($m_c=1.5\GeV$) and strange ($m_s=0.45\GeV$) quarks, respectively.}
\label{fig_ptquark}
\end{figure}
Fig.~\ref{fig_ptquark} summarizes the results of the Langevin
simulations for the $p_t$ probability distributions (integrated over
spatial coordinates) for $c$ and $s$ quarks, compared to (normalized)
blast-wave spectra for $T=180 \MeV$, using the same flow field as in the
Langevin simulation. The average surface-expansion velocity is
$\erw{\beta_0} \simeq 0.48c$. At low $p_t$ the spectra approach the
equilibrium limit as to be expected since the background medium for the
Langevin simulation is assumed to be fully equilibrated at all $p_t$, as
in a hydrodynamic calculation, and the stochastic process has been
realized as to guarantee the correct equilibrium limit (including the
adjustment of the (longitudinal) diffusion coefficient according to
Einstein's fluctuation-dissipation relation~\cite{vanHees:2005wb}). As
discussed above, the interaction strength has been chosen to recover the
empirically observed maximum elliptic flow of $v_{2^,q}^{\mathrm{max}}
\simeq 7$-$8\%$. Note, however, that the assumption of a fully
thermalized background medium implies rather large $v_2$ values at high
$p_t$, while the phase-space density of thermal partons is rather
small. A full treatment of this problem would require to solve a
selfconsistency problem, where the parton spectra in the background
fireball evolution also exhibit a saturation of the elliptic flow,
$v_2(p_t)$. This is beyond the scope of the present paper.

\section{Meson $p_T$ Spectra and $v_2$ Systematics}
\label{sec_meson}

We now combine the ingredients of our resonance recombination model
(RRM) by implementing the quark spectra computed from Langevin
simulations in the previous Section (\ref{sec_parton}) with the
Boltzmann-based hadronization formalism of Sec.~\ref{sec_boltz},
evaluated in the stationary (equilibrium) limit according to
Eq.~(\ref{eqboltzappr}). The key issue we address is how the properties
of the input (non-observable) quark spectra reflect themselves in the
(observable) meson spectra, in particular whether the space-momentum
correlations generated in the Langevin simulations can be consistent
with the empirically observed CQNS, which, in turn, opens a window on
the quark spectra at hadronization.

It remains to specify the masses and widths of hadrons in the
recombination process. In line with our restriction to mesons located
above the quark-antiquark threshold (due to the limitation to $2 \to 1$
processes) we consider $s$-$\bar{s}$, $c$-$\bar{c}$ coalescence with
resonance masses corresponding to the vacuum values for $\phi$
($1.02\,GeV$) and $J/\psi$ ($3.1\,GeV$) mesons; in connection with the
quark masses as given above this implies similar $Q$ values of
$0.1$-$0.12 \GeV$. The (total) meson widths are chosen of comparable
magnitude, \ie, $\Gamma_{\phi}=0.05 \GeV$ and
$\Gamma_{J/\psi}=0.1\,\mathrm{GeV}$. As elaborated in
Ref.~\cite{Ravagli:2007xx}, the numerical results, especially for the
meson $v_2$, are rather insensitive to variations in the meson width as
long as $Q$ is positive and substantially smaller (not smaller) than the
resonance mass (width).

\begin{figure}[!t]
\includegraphics[width=0.4\textwidth]{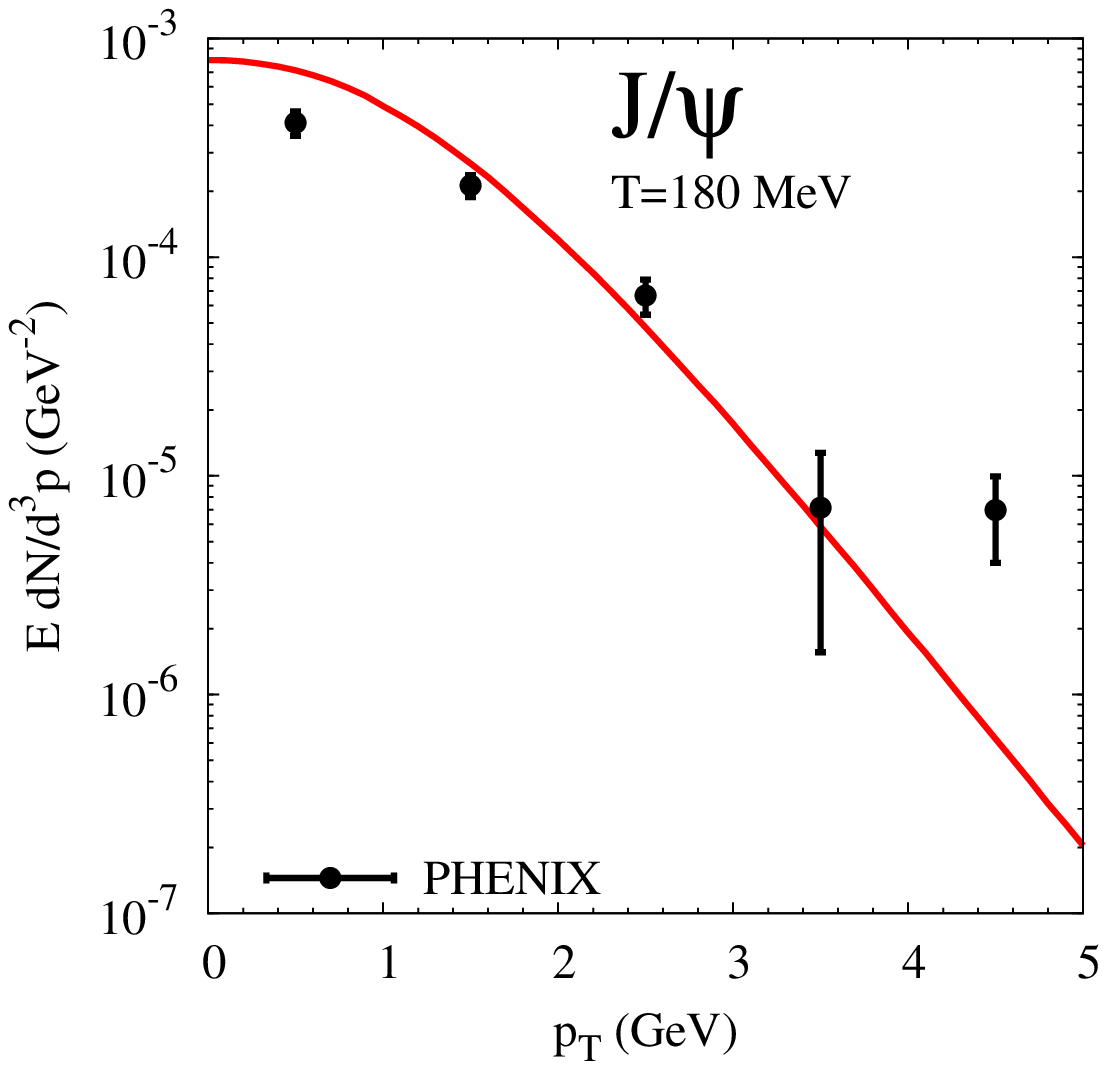} 
\hspace{1.0cm}
\includegraphics[width=0.4\textwidth]{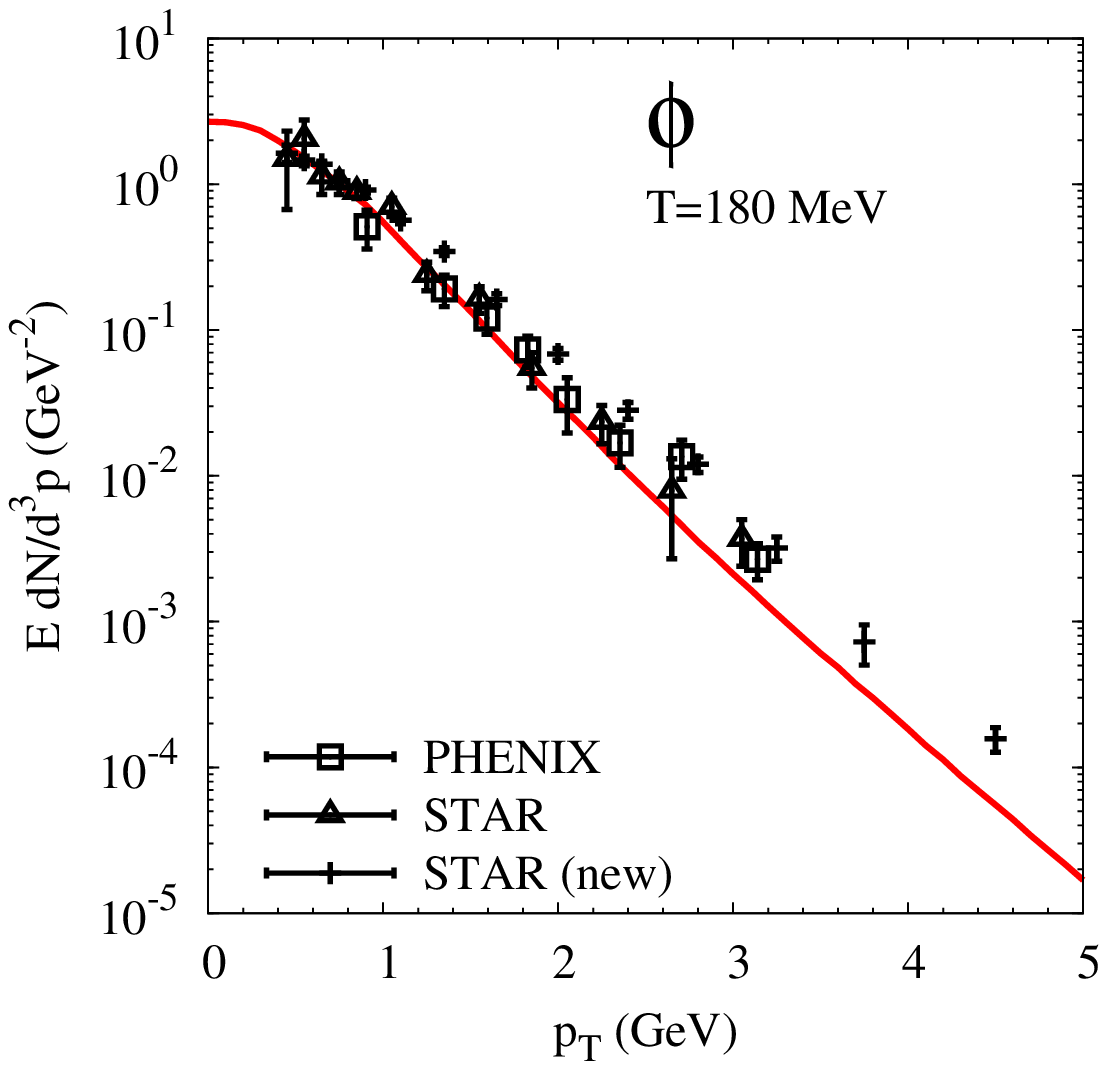}
\caption{(Color online) Meson $p_T$ spectra from quark-antiquark
  coalescence in central $\sqrt{s_{NN}}=200 \GeV$ Au-Au collisions
  computed within the resonance recombination model for $J/\psi$ (left
  panel) and $\phi$ (right panel); experimental data are from
  Refs.~\cite{Adare:2006ns} and
  \cite{Adler:2004hv,Adams:2004ux,Abelev:2007rw}, respectively.}
\label{ptmeson}
\end{figure}

In Fig.~\ref{ptmeson} we display our RRM results for $p_T$ spectra of
$J/\psi$ and $\phi$ mesons, including available RHIC data. Overall, the
spectra largely agree with those computed in our previous work (Fig.~1
in Ref.~\cite{Ravagli:2007xx}), where the input spectra were solely
based on a blast-wave parameterization. This is not surprising, since
the quark spectra employed in the present work show a rather large
degree of thermalization, up to momenta of $p_t=1.5$-$2 \GeV$ for
strange and charm quarks, cf.~Fig.~\ref{fig_ptquark}. Consequently, the
$\phi$-meson spectra shown here are somewhat harder than in
Ref.~\cite{Ravagli:2007xx} beyond $p_T\simeq 3\GeV$ due to the presence
of the kinetic (hard) components in the quark spectra resulting from the
Langevin evolution. The computed spectra for the $J/\psi$ are quite
reminiscent to earlier blast-wave based
results~\cite{Greco:2003vf,Andronic:2006ky,Zhao:2007hh}. Note, however,
that in Ref.~\cite{Zhao:2007hh} the recombination (blast wave)
contribution only amounts to about $50\%$ of the total $J/\psi$ yield,
significantly less than in the present paper (this is sensitive to the
total open-charm cross section, which is not very well determined yet).

\begin{figure}[!t]
\includegraphics[width=0.4\textwidth]{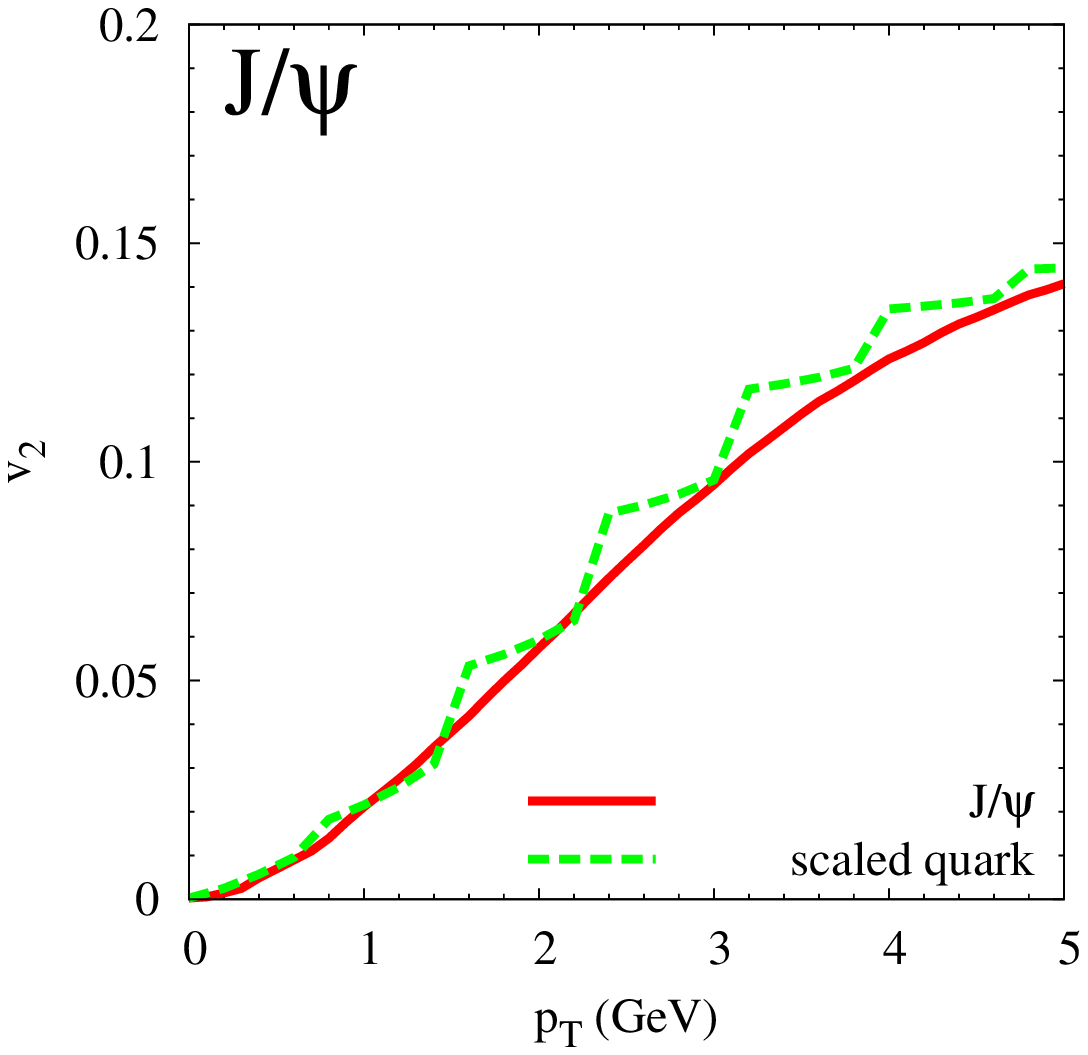}
\hspace{1.0cm}
\includegraphics[width=0.4\textwidth]{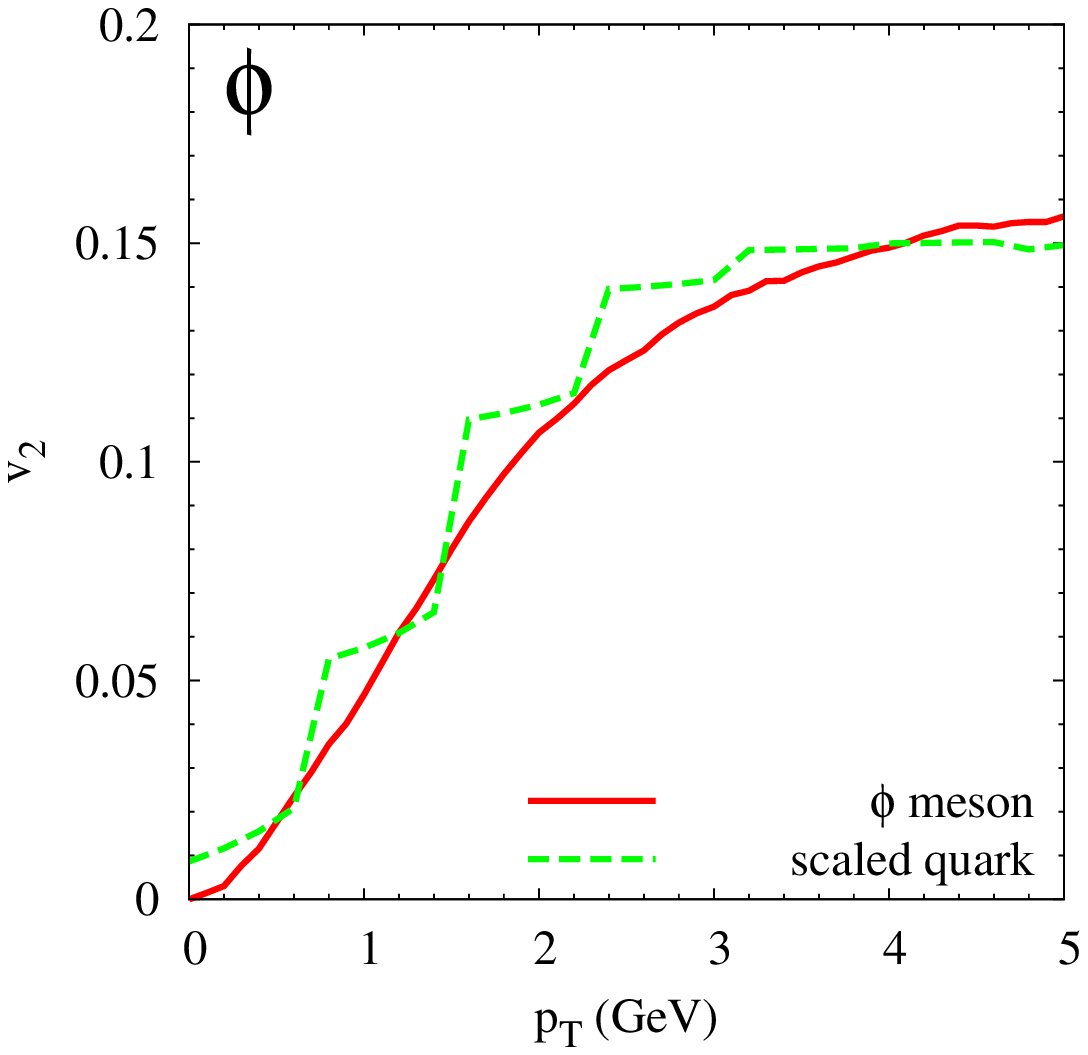}
\caption{(Color online) Scaled quark $v_2^{\rm scaled}$ (dashed lines)
  and meson $v_{2,M}$ (solid lines) coefficients as a function of meson
  $p_T$ for $J/\Psi$ ($Q=0.1\GeV$, $\Gamma=0.1\GeV$; left panel) and
  $\phi$ ($Q=0.12\GeV$, $\Gamma=0.05\GeV$; right panel) mesons.}
\label{v2meson}
\end{figure}

The RRM results for the meson $v_2(p_T)$ are summarized in
Fig.~\ref{v2meson} (solid lines) and compared to the underlying quark
$v_2$ \emph{scaled} to meson variables in the conventional (empirical)
way as $v_{2}^{\mathrm{scaled}}(p_T)\equiv 2v_{2,q}(p_T/2)$. We find
that for both the $J/\psi$ and $\phi$ the agreement is rather
impressive, within a few percent relative deviation. The wiggles at the
quark level are, to a large extent, driven by the finite grid sampling
due to a step width of $400 \MeV$ (the statistical error inherent to the
Langevin simulation is smaller than that, using $10^8$ test
particles). The convolution of two quark distributions results in much
smoother curves at the meson level (due to the fitting procedure of the
quark input). We are thus able to approximately recover CQNS in a
microscopic calculation with the full information on space-momentum
correlations, characteristic for hydrodynamic expansion at low $p_t$ and
a kinetic regime at higher $p_t$.

\begin{figure}[!t]
\includegraphics[width=0.4\textwidth]{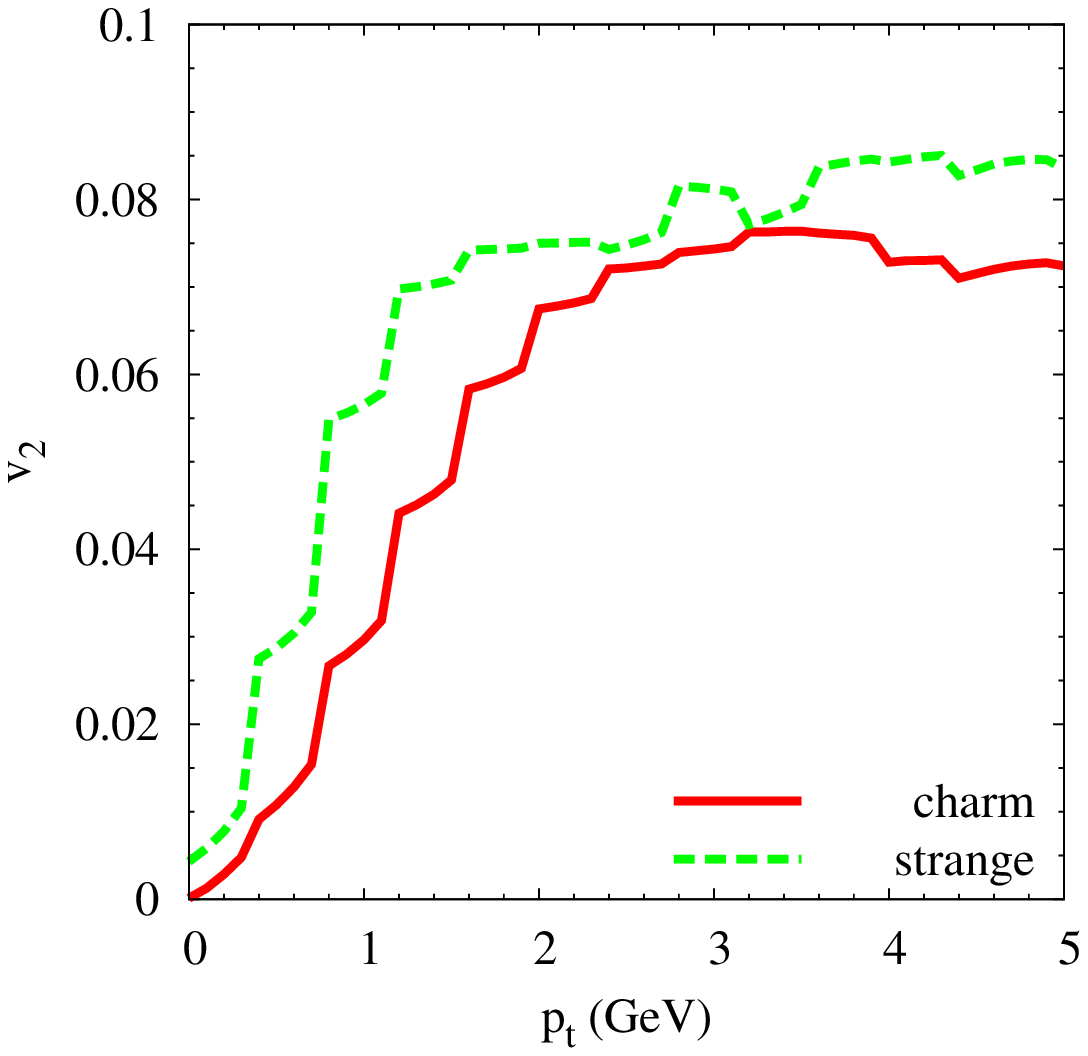}
\hspace{1.5cm}
\includegraphics[width=0.4\textwidth]{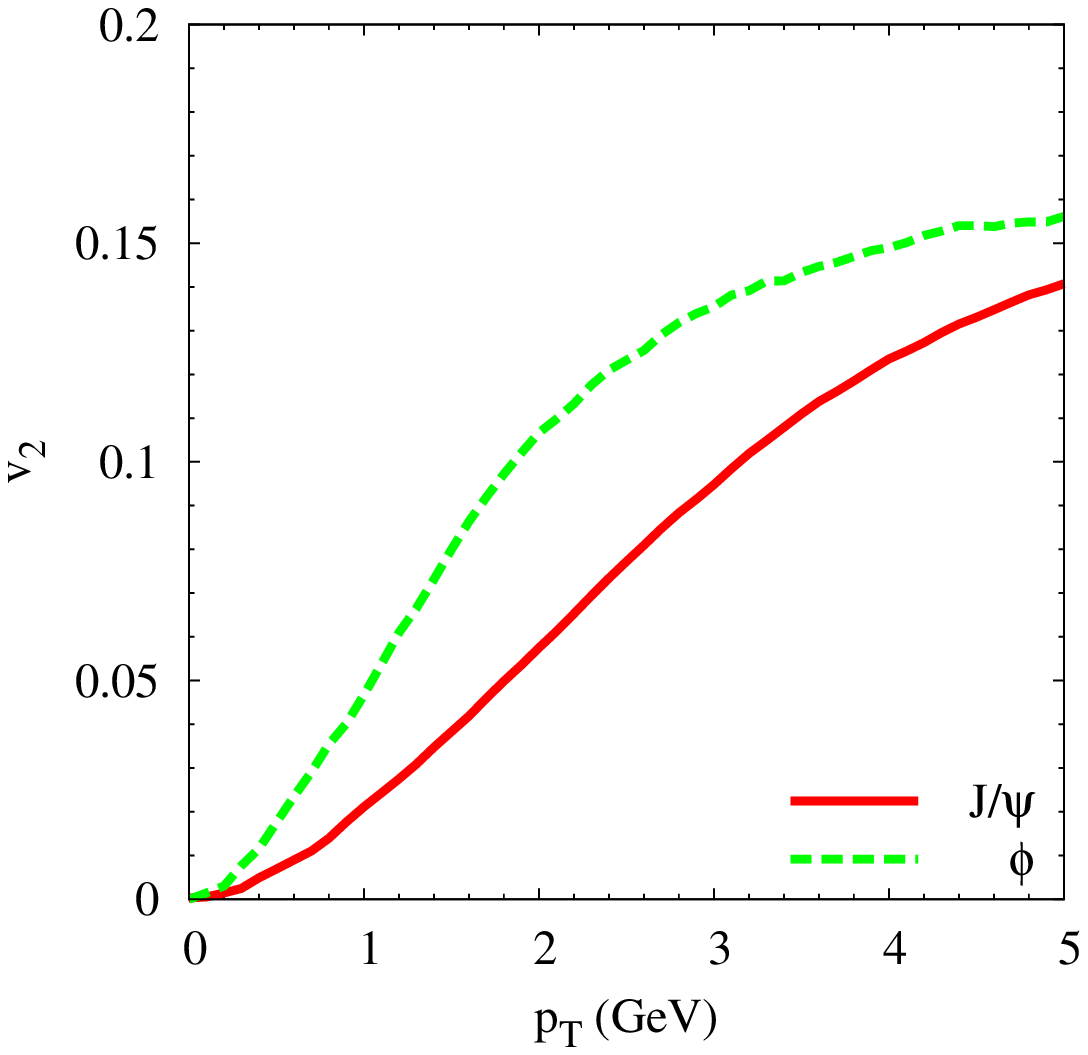} 
\caption{(Color online) Elliptic flow coefficient, $v_2$, as a function
  of transverse momentum for $c$ and $s$ quarks (left panel), as
  well as $J/\psi$ and $\phi$ mesons resulting from quark
  recombination (right panel), in semicentral $\sqrt{s_{NN}}=200\GeV$
  Au-Au collisions.}
\label{v2pT}
\end{figure}

Another potential source for scaling violations are flavor (or mass)
dependencies at the quark level (rather than in the coalescence
process). In Fig.~\ref{v2pT} we compare the elliptic flow of different
quarks (left panel) and mesons (right panel) with each other. The left
panel confirms that the $c$-quark $v_2$ deviates significantly (up to
$\sim$$20\%$ at low $p_t$) from the strange-quark $v_2$. At high $p_t$,
the strange-quark $v_2$ is a bit high, which, in principle, could be
readjusted by a somewhat reduced strength of the resonance interactions
in the Langevin simulations. At the meson level, the differences are
similar. We have verified that comparable deviations persist when
reducing the strange quark $v_2$.

\begin{center}
\begin{figure}[!t]
\includegraphics[width=0.4\textwidth]{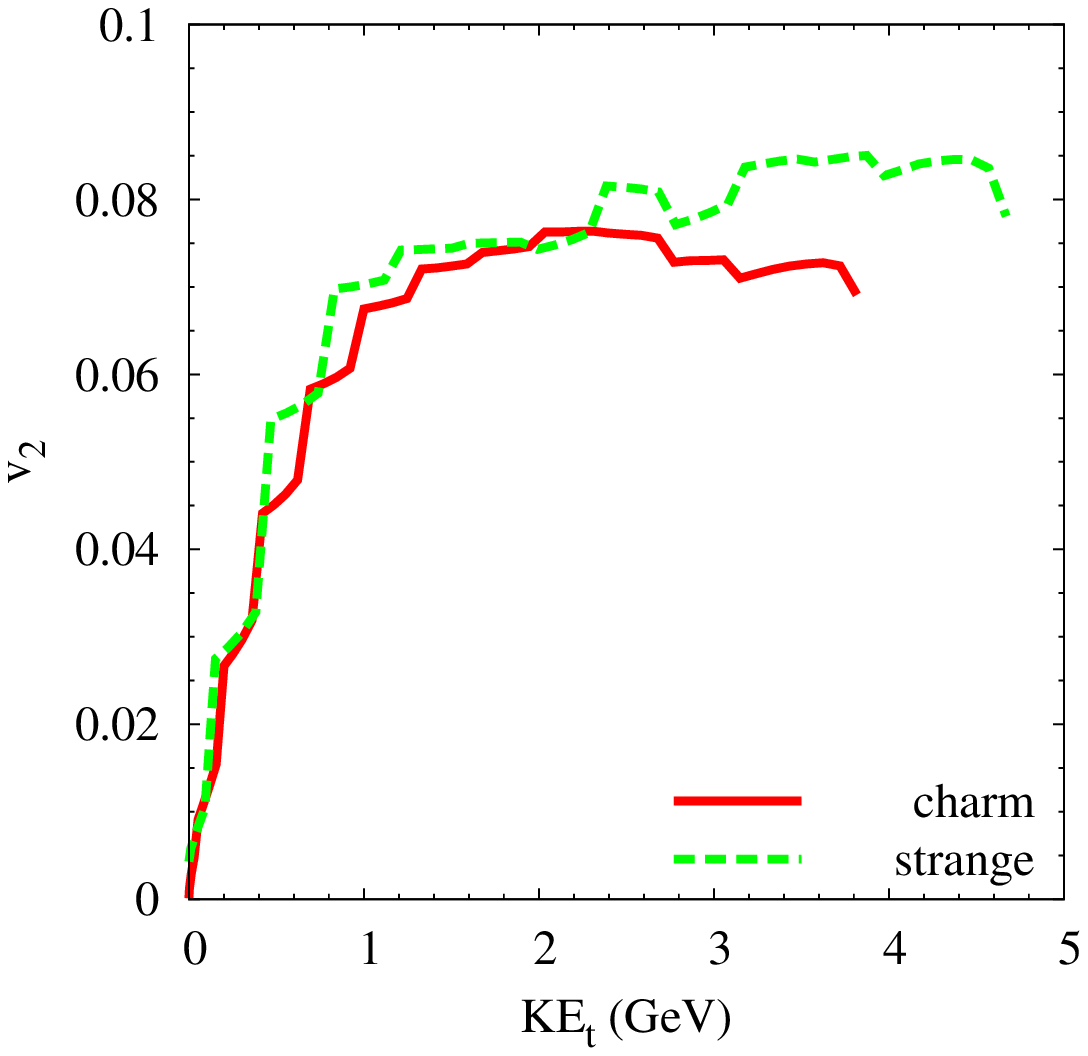}
\hspace{1.5cm}
\includegraphics[width=0.4\textwidth]{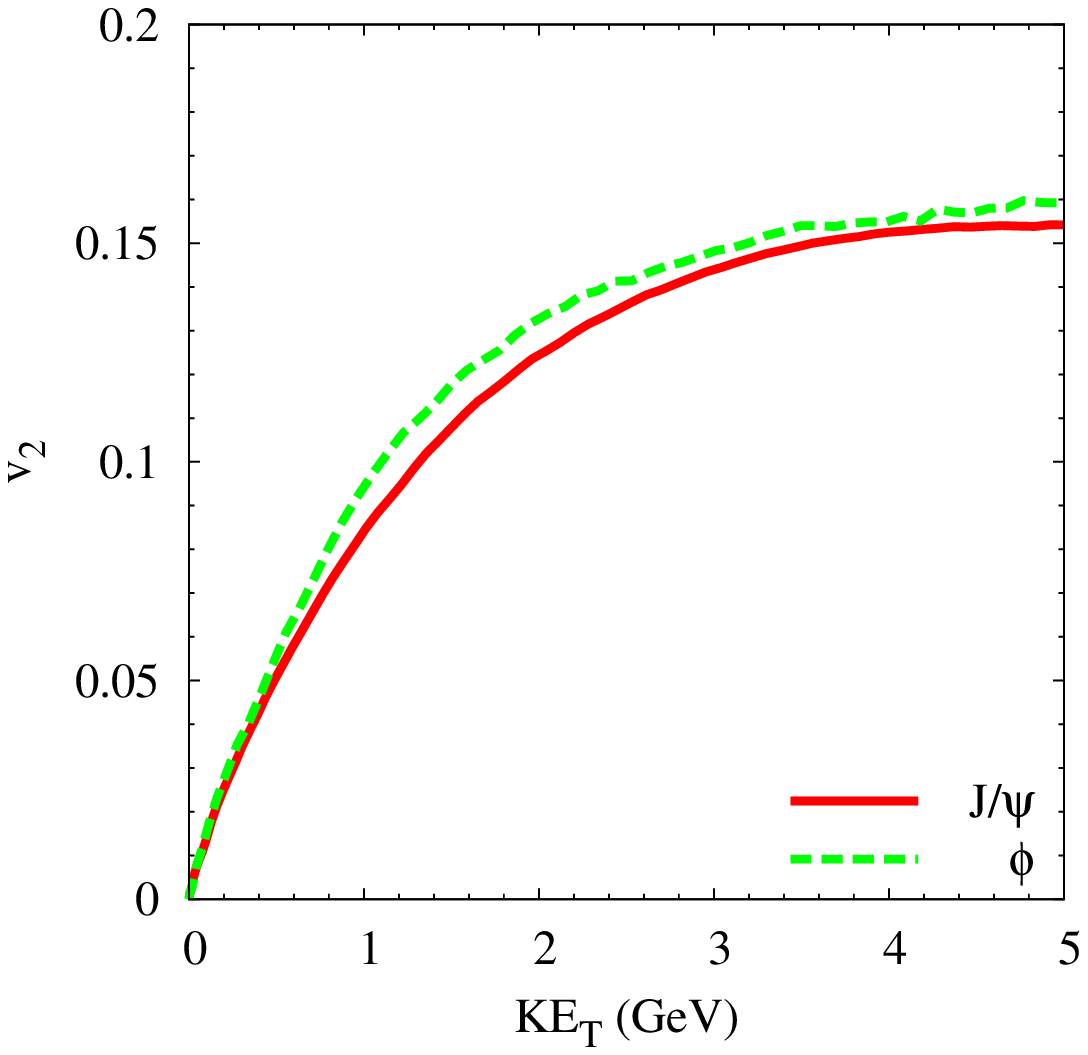} 
\caption{(Color online) Elliptic flow coefficient, $v_2$, as a function
  of transverse kinetic energy, $K_{t,T}$, for strange and charm quarks
  (left panel), as well as for $\phi$ and $J/\psi$ mesons (right
  panel).}
\label{v2KET}
\end{figure}
\end{center}

Finally, we address the question of CQNS with respect to transverse
kinetic energy, $K_T=m_T-m$, rather than transverse momentum, $p_T$.
Such a scaling has recently been highlighted by the
PHENIX~\cite{Adare:2006ti} and STAR collaborations~\cite{Abelev:2007qg}
and seems to be very well satisfied by all available RHIC data, in
centrality, collision energy and collision systems, after a geometric
correction for the nuclear overlap. $K_T$ scaling has also been found to
result from certain classes of hydrodynamic
solutions~\cite{Csanad:2005gv}, and therefore been argued to reflect a
collectively expanding thermalized system of
partons~\cite{Lacey:2006pn}. From the point of view of quark
coalescence, the problem of reconciling quark distribution functions
with space-momentum correlations (as implied by hydrodynamic expansion)
with CQNS persists. In Fig.~\ref{v2KET} we display the RRM results for
$v_{2,q}$ and $v_{2,M}$ for the two different flavors as a function of
$K_{t,T}$. We find that the quark input distributions from the QGP
Langevin simulations indeed share a rather universal behavior up to $K_t
\simeq3\,\mathrm{GeV}$, encompassing both the quasi-equilibrium regime
at low energies and the kinetic regime at intermediate energies
characterized by a leveling off at $K_t\gsim 1\GeV$, cf.~left panel of
Fig.~\ref{v2KET}. We recall that the only adjusted input to this result
is the common maximum value of the individual quark elliptic flow at
about $7$-$8\%$ (as suggested by the empirical CQNS deduced from
experiment), controlled by the nonperturbative interaction strength in
the stochastic Langevin force (again, a fine tuning for the $s$-quark
would improve the agreement at higher $K_t$). The approximate
universality at the quark level is nicely preserved at the meson level
as a result of our Boltzmann-based recombination formalism, see right
panel of Fig.~\ref{v2KET}. This is a quite remarkable result in view of
the underlying space-momentum correlations in our approach, which has
not been achieved before in this form.

\section{Conclusions}
\label{sec_concl}

In the present paper we have extended our previously formulated quark
coalescence formalism, utilizing resonance interactions within a
Boltzmann equation, by implementing microscopic quark phase-space
distributions generated via Langevin simulations of an expanding thermal
QGP fireball for Au-Au collisions at RHIC. In this way we could combine
the merits of our recombination approach (energy conservation and a
proper equilibrium limit) with those of realistic quark distributions,
which in particular encode the transition from a thermal regime at low
$p_t$ to a kinetic one at intermediate $p_t$. The latter feature is
especially important as it produces the leveling-off of the elliptic
flow, a key feature of observed hadron spectra and a crucial
prerequisite to test any kind of quark scaling behavior. The
(constituent) quark-mass and meson parameters were fixed at rather
standard values, and we have constrained ourselves to mesons which are
reliably calculable in our $2\to1$ recombination setup (\ie, for
positive $Q$ values). The only real adjustment concerned the interaction
strength in the Langevin process, as to reproduce the empirical maximum
value for the quark $v_2$ (with the QGP fireball parameters tuned to
empirical values of radial and elliptic flow, as in earlier applications
to, \eg, heavy-quark observables). These interactions have been modeled
via (meson) resonances in the QGP, which we identified with the states
formed in the coalescence process at $T_c$, leading to the notion of a
``Resonance Recombination Model'' (RRM). Since the Fokker-Planck
approach as an expansion of the Boltzmann equation is strictly valid for
sufficiently massive and/or high-momentum particles, we restricted
ourselves to ``heavy'' flavors, i.e., charm and strange quarks. At low
$p_t$, the latter are at the borderline of applicability of a
Fokker-Planck framework.

Our main finding is that within this rather generic set-up, largely
based on first principles augmented by a concrete realization of the
strongly interacting QGP, the constituent quark scaling of the meson
elliptic flow emerges rather naturally \emph{including} space-momentum
correlations characteristic for a collectively expanding source. The
scaling holds for individual mesons, but appears to be rather universal
in quark and meson flavor (mass), especially when applied in
(transverse) kinetic energy rather than momentum, which is in line with
recent experimental findings. By overcoming some of the limitations of
previous (more schematic) coalescence models, and by achieving the first
robust implementation of realistic (microscopically computed)
phase-space distribution functions of quarks, our formalism could
provide a useful tool to better understand systematics of RHIC data,
most notably the interplay of a thermal and kinetic regime in connection
with phase-space properties of the partonic fireball as viewed through
the hadronization process.

Clearly, a formidable list of open issues persists, including the
extension to light quarks, the role of gluons and of deeply bound
hadronic states (possibly requiring additional formation processes),
more realistic spectral functions of mesons and quarks and their
interactions (both around $T_c$ and above), a selfconsistent treatment
of thermal and kinetic components (possibly requiring full parton
transport), a systematic classification of viable parton phase-space
distributions, hadronic reinteractions, etc.  Progress has already been
made on a number of these aspects, but a comprehensive approach remains
a challenging task.


\begin{acknowledgments}
  We thank T.~Hahn for insightful clarifications about the CUBA
  multi-dimensional integration package~\cite{Hahn:2004fe} which has
  been used in this work, and R.~J.~Fries for valuable discussions. This
  work was supported in part by a U.S. National Science Foundation
  CAREER award under grant no. PHY-0449489 and by the A.-v.-Humboldt
  Foundation (through a Bessel Research Award).
\end{acknowledgments}


\begin{flushleft}

\end{flushleft}

\end{document}